\documentclass[11pt]{article}

\usepackage{epsfig}
\usepackage{graphicx}
\usepackage{epstopdf}
\usepackage{amsfonts}
\usepackage{amssymb}
\usepackage{amsthm}
\usepackage{amsmath}
\usepackage{multirow}
\usepackage{color}

\def\beq{\begin{equation}}
\def\eeq{\end{equation}}
\def\bea{\begin{eqnarray}}
\def\eea{\end{eqnarray}}
\newcommand{\beqs}{\begin{subequations}}
\newcommand{\eeqs}{\end{subequations}}

\newcommand{\cref}[1]{Ref.~\cite{#1}}

\newcommand{\vev}[1]{\left<#1\right>}

\newcommand{\Tr}{\mathsf{Tr}}

\newcommand{\hh}{{\ensuremath{I{\kern-2.6pt h}}}}
\newcommand{\bhh}{{\ensuremath{\bar{I{\kern-2.6pt h}}}}}

\begin{document}

\begin{titlepage}
	
\vspace*{-15mm}
\begin{flushright}
{UT-STPD-21/01}\\
\end{flushright}
\vspace*{0.7cm}

\begin{center}
{\Large {\bf Triply Charged Monopole and Magnetic Quarks
}}
\\[12mm]
George Lazarides,$^{1}$~%\footnote{E-mail: \texttt{lazaride@eng.auth.gr}}
Qaisar Shafi$^{2}$~%\footnote{E-mail: \texttt{shafi@bartol.udel.edu}}
\end{center}
\vspace*{0.50cm}
\centerline{$^{1}$ \it
School of Electrical and
Computer Engineering, Faculty of Engineering,
}
\centerline{\it
Aristotle University
of Thessaloniki, Thessaloniki 54124, Greece}
	\vspace*{0.2cm}
	\centerline{$^{2}$ \it
		Bartol Research Institute, Department of Physics and 
		Astronomy,}
	\centerline{\it
		 University of Delaware, Newark, DE 19716, USA}
	\vspace*{1.20cm}
\begin{abstract}
We describe the internal composition of a topologically 
stable monopole carrying a magnetic charge of $6\pi/e$ 
that arises from the spontaneous breaking of the 
trinification symmetry $SU(3)_c\times SU(3)_L\times 
SU(3)_R$ ($G$). Since this monopole carries no color 
magnetic charge, a charge of $6\pi/e$ is required by 
the Dirac quantization condition. The breaking  of $G$ 
to the Standard Model occurs in a number of steps and 
yields  the desired topologically stable monopole 
(``magnetic baryon''), consisting of three confined 
monopoles. The confined monopoles (``magnetic quarks'') 
each carry a combination of Coulomb magnetic flux and 
magnetic flux tubes, and therefore they do not 
exist as isolated states. We also display a more 
elaborate configuration (``fang necklace'') composed of
these magnetic quarks. In contrast to the 
$SU(5)$ monopole which is superheavy and carries a magnetic 
charge of $2\pi/e$ as well as color magnetic charge, 
the trinification monopole may have mass in the TeV 
range, in which case it may be accessible at the LHC and 
its planned upgrades. 
\end{abstract}

\vspace{1.5cm}
{\Large\textit{\textbf{\textsl{
\begin{center}
The Magnetic Monopole Ninety Years Later
\end{center}
}}}}
\end{titlepage}

\section{Introduction}
Grand unification theories (GUTs) based on a single gauge 
coupling such as $SU(5)$ \cite{GeorgiGlashow} predict the 
existence of a topologically stable magnetic monopole which 
carries one quantum ($2\pi/e$) of Dirac magnetic charge 
\cite{dokos,daniel}. In contrast to the 't Hooft-Polyakov 
monopole \cite{monopole}, the $SU(5)$ monopole also carries 
an appropriate amount of color magnetic flux  
that is screened because of color electric confinement. 

Unification models based on product groups such as $SU(4)_c
\times SU(2)_L\times SU(2)_R$ \cite{pati} predict the 
existence of a topologically stable monopole that carries 
two quanta ($4\pi/e$) of magnetic charge \cite{magg}. One 
straightforward way to see this is by noting that the 
underlying group allows the existence of color singlet states 
that carry electric charges $\pm e/2$ and colored triplets 
with charges $\pm e/6$. A more explicit realization of this 
doubly charged monopole was demonstrated in Ref.~\cite{TopDef}, 
where it was shown to arise from the merger of two distinct 
(``confined'') monopoles, with each one carrying some Coulomb 
flux and a magnetic flux tube. This demonstration also reveals 
the existence of ``magnetic dumbbells'' in a variety of unified 
theories. 

Very interestingly, following Ref.~\cite{TopDef}, Volovik has 
shown \cite{volovik} how topological structures similar to the 
doubly charged construction in Ref.~\cite{TopDef} may arise in 
superfluid $^3{\rm He}$. Furthermore, the existence of a class 
of topological structures called ``walls bounded by strings'' 
\cite{kibble} was verified in experiments with superfluid 
$^3{\rm He}$ \cite{volovik2019}. Motivated by these recent 
developments and especially the interplay between topological 
structures in high energy and condensed matter physics, we 
explore some interesting topological structures that arise in 
the framework of the trinification gauge symmetry $G=SU(3)_c
\times SU(3)_L\times SU(3)_R$ \cite{TopDef,kephart}. In contrast 
to $SU(5)$ and $SU(4)_c\times SU(2)_L\times SU(2)_R$, the 
topologically stable monopole in the trinification model is 
purely electromagnetic in nature with no color magnetic field 
accompanying it. It carries three quanta of magnetic charge 
($6\pi/e$) in order to satisfy the Dirac quantization condition, 
and its mass may be light enough to make it accessible in high 
energy colliders. To identify the variety of topological 
substructures potentially associated with this monopole, we 
assume that the trinification symmetry breaking to the Standard 
Model (SM) proceeds 
through a series of steps. This deconstruction procedure allows us 
to identify the building blocks that make up the triply charged 
monopole. The latter, it turns out, consists of three distinct 
constituent monopoles which are bound together by flux tubes. We 
may thus refer to the triply charged monopole as a 
``magnetic baryon,'' and its confined constituent components as 
``magnetic quarks.'' It is clear that other bound states such as 
``magnetic mesons'' are also present in this trinification model. 
We display an example of a somewhat more elaborate topological 
configuration referred to as a ``fang necklace.''

\section{Triply Charged Monopole}

The trinification symmetry $G$ is a well known subgroup of $E_6$ 
\cite{ramond}, and a variety of topological structures that arise 
when the latter breaks to the SM have been discussed 
in Ref.~\cite{TopDef}. In this paper we do not insist on this 
relationship between the two groups, which allows us to 
contemplate the spontaneous breaking of $G$ at scales lying in 
the TeV range. Because $G$ implements electric charge conservation, 
its spontaneous breaking to the SM and subsequently to 
$SU(3)_c\times U(1)_{em}$ yields a topologically stable magnetic 
monopole that carries three quanta of Dirac magnetic charge, namely 
$6\pi/e$ \cite{kephart}.

Recall that in the presence of fractionally charged quarks, say $d$ 
or $s$, one naively expects the magnetic monopole to carry this 
amount ($6\pi/e$) of magnetic charge from the Dirac quantization 
condition ($\mathsf{q}\mathsf{g}/4\pi=n/2$, where $\mathsf{q}$, 
$\mathsf{g}$ denote the electric and magnetic charges respectively
and $n$ is an integer) \cite{dirac}. However, the topologically 
stable magnetic monopole in 
$SU(5)$ carries just a single quantum ($2\pi/e$) of magnetic 
charge. This is compatible with the Dirac quantization condition 
because the monopole also carries an appropriate amount of color 
magnetic charge \cite{daniel}. In the trinification case this is 
not the case and so the magnetic charge carried by the monopole is 
$6\pi/e$. A simple way to see this is to note that $G$ allows, in 
principle, color singlet states in the representations 
${\bf (1,3,1)}$ + h.c., which carry electric charge $\pm 1/3$, and 
therefore the magnetic monopole must carry a magnetic charge of 
$6\pi/e$. (Fractionally charged color singlet states accompanied 
by multiply charged monopoles also appear in string theories 
\cite{WenWitten}.) Recall that the observed 
quarks and leptons reside in bifundamental representations of $G$ 
such as ${\bf (1, \bar{3}, 3)}$, etc. The discussion regarding the 
monopole charge is a bit more subtle if $G$ is embedded in $E_6$ 
but the outcome remains intact \cite{TopDef,kephart}. The monopole
is topologically stable because the second homotopy group of the 
vacuum manifold $\pi_2(G/H)=\mathbb{Z}=\{n=0,\pm 1,\pm2, \pm 3,
...\}$, with $G$ being the trinification group and $H=SU(3)_c
\times U(1)_{em}$.

We now turn to the breaking of $G$ to $SU(3)_c\times 
SU(2)_L\times U(1)_{Y_L}\times SU(2)_R\times U(1)_{Y_R}$ 
at an intermediate scale, which can approach the TeV scale 
if desired. This is achieved by the vacuum expectation 
values (VEVs) of the 
{\bf (1,8,1)} and {\bf (1,1,8)} components of a Higgs 
{\bf 78}-plet under $E_6$. (It is sometimes convenient to 
follow the $E_6$ notation.) Recall that the $SU(3)_{L(R)}$ 
octet under the $SU(2)_{L(R)}\times U(1)_{Y_L(Y_R)}$ 
subgroup is decomposed as follows ${\bf 8=1_0+3_0+2_3+
2_{-3}}$, where the subscripts denote the charges with 
respect to the generator $T^8_{L(R)}\equiv 
{\rm diag}(1,1,-2)$ of $U(1)_{Y_L(Y_R)}$. We can further 
break $SU(2)_R$ to $U(1)_R$ by a VEV along the ${\bf 3_0}$ 
component of the $SU(3)_R$ octet. The generator of 
$U(1)_R$ is $T^3_R={\rm diag}(1,-1)$. The unbroken subgroup
is then $SU(3)_c\times SU(2)_L\times U(1)_{Y_L}\times 
U(1)_R\times U(1)_{Y_R}$.

To be a bit more explicit, the potential for the breaking 
of $SU(3)_R$ to $SU(2)_R\times U(1)_{Y_R}$, assuming a 
discrete symmetry $\phi\to -\phi$, with $\phi$ being the 
scalar octet, is given by
\begin{equation}
V=-\frac{1}{2}m^2\Tr\phi^2+\frac{a}{4}\Tr(\phi^2)^2
+ \frac{b}{2}\Tr\phi^4,
\end{equation} 
where $m$ is a mass parameter and $a$, $b$ dimensionless
parameters. The $3\times 3$ matrix $\phi_i^j$ can be 
diagonalized by an $SU(3)_R$ rotation, and for suitable 
choices of $a$ and $b$, $\phi$ acquires a VEV 
\begin{equation}
\vev{\phi}\propto
\begin{pmatrix}
1 & 0 & 0
\\
0 & 1 & 0 
\\
0  & 0 & -2
\end{pmatrix},
\end{equation}
which breaks $SU(3)_R$ to $SU(2)_R\times U(1)_{Y_R}$. With a 
second scalar octet, it is then straightforward to break 
$SU(2)_R$ to $U(1)_{R}$. More details will not be provided 
here. 

At this stage, we have the generation of three types of 
intermediate scale magnetic monopoles.
Two of them result from the breaking of $SU(3)_L$ and 
$SU(3)_R$ to $SU(2)_L\times U(1)_{Y_L}$ and $SU(2)_R
\times U(1)_{Y_R}$ and carry one unit of Coulomb magnetic 
flux along the generators $T^3_L/2+T^8_L/2$ and 
$T^3_R/2+T^8_R/2$ respectively, where $T^3_{L(R)}\equiv 
{\rm diag}(1,-1)$. This is because the $(-1,-1)\in 
SU(2)_{L(R)}\times U(1)_{Y_L(Y_R)}$ coincides with the 
identity element as it leaves all the representations of 
$SU(3)_{L(R)}$ unchanged. Consequently, a rotation by 
$2\pi$ along the generator $T^3_{L(R)}/2+T^8_{L(R)}/2$, 
which interpolates between (1,1) and (-1,-1), is a closed 
loop generating the second homotopy group $\pi_2
(SU(3)_{L(R)}/SU(2)_{L(R)}\times U(1)_{Y_L(Y_R})=\pi_1
(SU(2)_{L(R)}\times U(1)_{Y_L(Y_R)})=\mathbb{Z}$ of the 
vacuum manifold. The breaking of $SU(2)_R$ to $U(1)_R$ 
generates a third monopole which carries one unit of $T^3_R$ 
magnetic flux corresponding to a $2\pi$ rotation along this 
generator.

We should further break $U(1)_{Y_L} \times U(1)_R\times 
U(1)_{Y_R}$ to $U(1)_Y$, where $Y=T^3_R/2+(T^8_L+T^8_R)/6$ 
is the weak hypercharge. (The electric charge operator is 
given by $Q=T^3_L/2+Y$.) First consider the breaking of
$U(1)_{Y_L} \times U(1)_{Y_R}$ to $U(1)_{B-L}$, where 
$B-L=(T^8_L+T^8_R)/3$ is the baryon minus lepton number. 
This symmetry breaking is achieved by a Higgs field in the 
fundamental representation of $E_6$,  
\begin{equation}
{\bf 27}={\bf (1,\bar{3},3)}+{\bf (3,3,1)}+
{\bf (\bar{3},1,\bar{3})}\equiv\lambda+\mathbb{Q}+
\mathbb{Q}^c, 
\end{equation}
where
\begin{equation}
\lambda= 
\begin{pmatrix}
h_u & e^c
\\
&
\\
h_d & \nu^c 
\\
&
\\  
l  & N
\end{pmatrix}        
\end{equation}         
with the rows being ${\bf \bar{3}}$'s of $SU(3)_L$ and the 
columns {\bf 3}'s of $SU(3)_R$, and
\begin{equation}
\mathbb{Q}= 
\begin{pmatrix}
q
\\
&
\\
g
\end{pmatrix} \quad {\rm and}\quad
\mathbb{Q}^c=
\begin{pmatrix}
u^c, &d^c, &g^c 
\end{pmatrix},   
\end{equation}
denote an $SU(3)_L$ triplet and an $SU(3)_R$ antitriplet 
respectively. For simplicity, we use 
here for the various components of the Higgs {\bf 27}-plet 
the same symbols as for the corresponding components of the 
fermion {\bf 27}-plets which contain the ordinary quarks and 
leptons. The reader should keep this in mind to avoid any
confusion. The Higgs {\bf 27}-plet acquires a VEV along its 
$N$ component which is an $SU(3)_c\times SU(2)_L\times SU(2)_R$ 
singlet and has $T^8_L=2$, $T^8_R=-2$. Consequently, the
generator $T^8_L+T^8_R=3(B-L)$ remains unbroken \cite{TopDef}. 
A rotation by $2\pi/4$ along the orthogonal broken generator 
\beq
\mathcal{B}\equiv T^8_L-T^8_R
\eeq   
leaves the VEV of $N$ invariant. 
Consequently, the cosmic string generated by the breaking of 
$U(1)_\mathcal{B}$ is a tube with magnetic flux corresponding 
to this rotation, namely it carries magnetic flux 
$(T^8_L-T^8_R)/4$.   

We next consider the breaking $U(1)_R\times U(1)_{B-L}$ to 
$U(1)_Y$, where $Y=T^3_R/2+(B-L)/2$ is the SM weak hypercharge, 
by a VEV along the $\nu^c$ component of the Higgs 
{\bf 27}-plet which has $T^3_R=-1$ and $B-L=1$. The normalized 
generators corresponding to $T^3_R$ and $(B-L)$ are $T^3_R/2$ 
and $\sqrt{3/8}(B-L)$ and, thus, the orthogonal broken generator 
is 
\beq
2T^3_R-3(B-L).
\label{orthgen1}
\eeq 
This generator is unbroken by the VEV of $N$, but breaks by the 
VEV of $\nu^c$. However, the charges of $\nu^c$ imply that a 
rotation by $2\pi/5$ along this generator remains unbroken and 
the associated string carries magnetic flux 
\beq
\frac{2}{5}T^3_R-\frac{3}{5}(B-L).
\label{tube}
\eeq

Revisiting the tube with magnetic flux $(T^8_L-T^8_R)/4$, we 
see that as we go around it the VEV of $\nu^c$ acquires a factor
$\exp(2i\pi/4)$ since its relevant charges are $T^8_L=2$, 
$T^8_R=1$. To cancel this factor, we should add along the tube 
an additional magnetic flux $(1/4)\{2T^3_R/5-3(B-L)/5\}$ so that 
$\nu^c$ acquires an extra factor $\exp(-2i\pi/4)$. This additional 
flux does not affect the VEV of $N$ since its relevant charges are
$T^3_R=0$, $B-L=0$. In conclusion, we obtain a tube with a 
combined magnetic flux 
\beq
\frac{1}{4}(T^8_L-T^8_R)+\frac{1}{4}\{\frac{2}{5}T^3_R-
\frac{3}{5}(B-L)\}.
\label{combined}
\eeq

In Ref.~\cite{TopDef}, it has been shown that the only intermediate 
scale topological defect which survives in this model, where the 
symmetry breaking employs the $\nu^c$ component of a Higgs 
${\bf 27}$-plet rather than the $\nu^c\nu^c$ component of a Higgs 
$\overline{{\bf 351}'}$, is a triply charged ($6\pi/e$) magnetic 
monopole. 
Therefore, one expects that the three types of intermediate scale 
monopoles and the two types of magnetic flux tubes mentioned above 
must combine to generate this monopole. Indeed, when the 
trinification group  is 
broken to the SM gauge group, the magnetic flux $T^3_R/2+T^8_R/2$ 
emerging from the $SU(3)_R$ monopole splits into two parts, one 
equal to minus the flux in Eq.~(\ref{combined}) which forms a tube 
and one Coulomb flux equal to $6Y/5$. Similarly, the magnetic flux 
$T^3_R$ of the $SU(2)_R$ monopole forms a tube with flux given 
in Eq.~(\ref{tube}) and a Coulomb magnetic field with flux $6Y/5$. 
This tube is absorbed by an $SU(3)_L$ monopole with flux $T^3_L/2
+T^8_L/2$, which also emits the tube with magnetic flux as 
in Eq.~(\ref{combined}) terminating on the $SU(3)_R$ monopole. The 
remaining magnetic flux $T^3_L/2+3Y/5$ forms a Coulomb magnetic 
field emerging from the $SU(3)_L$ monopole. At this point, it is 
convenient -- for reason to become apparent in the next paragraph 
-- to add to the Coulomb fields of the $SU(3)_R$ and the $SU(2)_R$ 
monopoles and subtract from the magnetic field of the $SU(3)_L$ 
monopole a magnetic flux $T^3_L$. This is legitimate since a 
rotation by $2\pi$ around $T^3_L$ is homotopically trivial. The 
sum of the Coulomb magnetic fluxes emerging from the three 
monopoles is then
\beq
\frac{3}{2}T^3_L+3Y=3Q,
\eeq        
where $Q$ is the electric charge operator. Consequently, the
three constituent magnetic monopoles (magnetic quarks) are pulled 
together by the strings to create a triply charged ($6\pi/e$) 
magnetic monopole. 

\begin{figure}[t]
\centerline{\epsfig{file=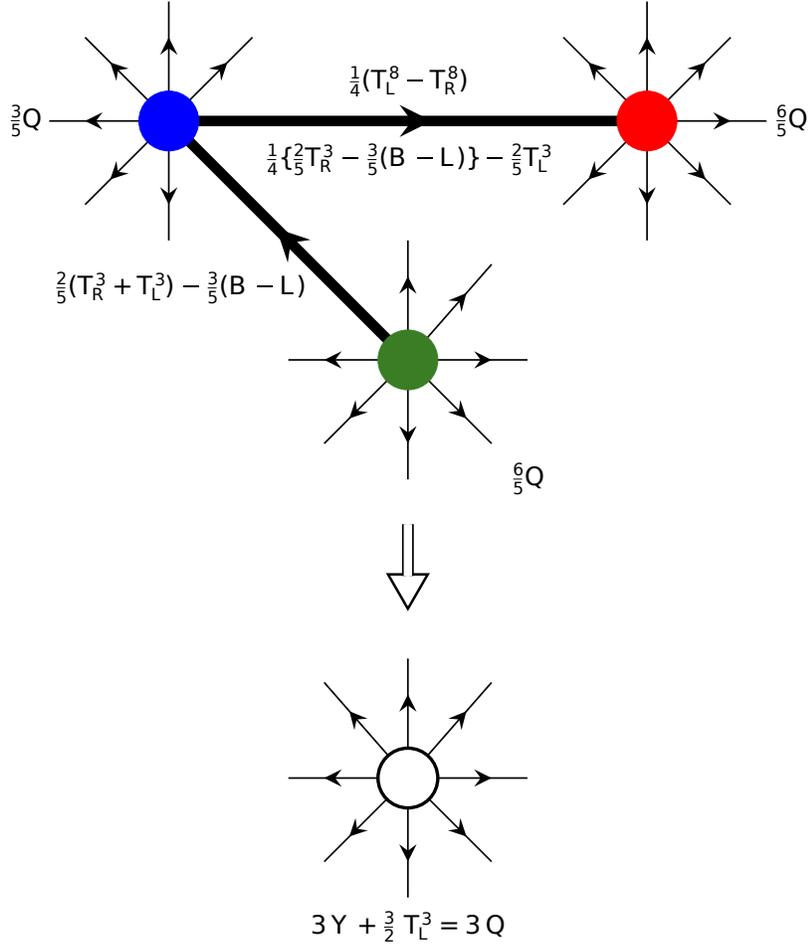,width=10.6cm}}
\caption{Emergence of the topologically stable triply 
charged monopole from the symmetry breaking 
$G
%=SU(3)_c\times SU(3)_L\times SU(3)_R
\to SU(3)_c\times 
SU(2)_L\times U(1)_{Y_L}\times U(1)_{Y_R}\times U(1)_R
\to SU(3)_c\times SU(2)_L\times U(1)_Y\to SU(3)_c\times 
U(1)_{em}$. An $SU(2)_R$ (green) monopole is connected 
by a flux tube to an $SU(3)_L$ (blue) monopole which, 
in turn, is connected to an $SU(3)_R$ (red) monopole by 
a superconducting flux tube. The constituent monopoles 
are pulled together to form the triply charged monopole. 
The fluxes along the tubes and around the monopoles are 
indicated.}
\label{fig:TripMon}
\end{figure}

Next we consider the effect of the electroweak symmetry breaking
on the two tubes with magnetic fluxes given in Eqs.~(\ref{tube}) 
and (\ref{combined}). The relevant charges of the VEVs $\vev{h_u}$ 
and $\vev{h_d}$ of the electroweak doublets $h_u$ and $h_d$, which 
couple to the up-type and down-type quarks, are $T^3_L=-1$, 
$T^3_R=1$, $T^8_L=-1$, $T^8_R=1$, and $T^3_L=1$, $T^3_R=-1$, 
$T^8_L=-1$, $T^8_R=1$, respectively. Consequently, as we go
around the string with magnetic flux as in Eq.~(\ref{combined}),
the phase of $\vev{h_u}$ changes by $(-2/5)2\pi$ and that of 
$\vev{h_d}$ by $(-3/5)2\pi$. The tube must then acquire an extra 
magnetic flux $-2T^3_L/5$ so that the phase of $\vev{h_d}$ 
changes by $-2\pi$ and $\vev{h_u}$ remains constant around the 
string. Similarly, as we go around the string with magnetic flux 
as in Eq.~(\ref{tube}), the phases of $\vev{h_u}$ and $\vev{h_d}$ 
change by $(2/5)2\pi$ and $(-2/5)2\pi$ respectively. Thus, we must 
add an extra magnetic flux $2T^3_L/5$ along this tube so that both 
$\vev{h_u}$ and $\vev{h_d}$ remain
constant around the string. This choice is energetically favored
since it minimizes the magnetic energy along the strings -- see 
Ref.~\cite{TopDef}. The Coulomb magnetic fluxes 
emerging from the $SU(3)_R$ and $SU(2)_R$ monopoles are 
$(6/5)(T^3_L/2+Y)=6Q/5$ each, and from the $SU(3)_L$ monopole this 
flux is equal to $(3/5)(T^3_L/2+Y)=3Q/5$, in total $3Q$ -- 
see Fig.~\ref{fig:TripMon}. 

The Coulomb magnetic charges accompanying the $SU(3)_R$, 
$SU(3)_L$, and $SU(2)_R$ constituent magnetic monopoles are, 
respectively, 
$(6/5)2\pi/e$, $(3/5)2\pi/e$, and $(6/5)2\pi/e$. These magnetic 
charges, by construction, are compatible with the Dirac 
quantization condition because of their accompanying magnetic 
flux tubes. (Magnetic monopoles carrying a mixture of Coulomb 
magnetic flux and $Z$-magnetic flux have been considered in the 
past \cite{magg,Ztube}. For a recent discussion see 
Refs.~\cite{TopDef,hung}.)

Clearly, each of the three types of constituent magnetic monopoles 
(magnetic quarks) can alternatively be connected to its own magnetic 
antiquark by the appropriate flux tube(s) to produce a magnetic 
meson in the case of the $SU(2)_R$ and $SU(3)_R$ monopoles with a 
single flux tube connecting it to its antimonopole, or a new type of
magnetic mesons in the case of the $SU(3)_L$ magnetic quark with two 
flux tubes connecting it to its magnetic antiquark. In all three 
cases, the magnetic quarks and antiquarks eventually annihilate by 
being pulled together. 

Let us briefly discuss the mass of the triply charged magnetic 
monopole. This mass depends, of course, on the breaking scale 
$M$ of the trinification symmetry. Since the latter is not a 
grand unified theory without additional assumptions 
such as gauge coupling constant unification, there is 
nothing, in principle, that prevents the scale $M$ to lie 
in the TeV range, in which case the magnetic monopole 
mass also is of order $M$ or somewhat larger. This would 
make the topologically stable trinification monopole 
accessible at the LHC \cite{moedal} and its planned 
upgrades. For completeness, let us note that the size of 
the core of each magnetic monopole is determined by 
$g M^{-1}$, where $g$ and $M$ denote the relevant gauge 
coupling constant and symmetry breaking scale. Also, the 
mass per unit length of the magnetic flux tubes is of 
order $\mu^2$, with $\mu$ being the corresponding symmetry 
breaking scale. These flux tubes are practically stable with
a relatively small hierarchy between $M$ and $\mu$.

Finally, some remarks regarding the observability of this 
topologically stable triply charged monopole at the LHC are in 
order here. It has been recognized for quite some time now that 
the production cross section of a composite coherent quantum state 
such as this monopole is expected to be exponentially suppressed 
in Drell-Yan processes involving elementary particles -- for a 
recent review and additional references, see Ref.~\cite{DrellYan}. 
This is somewhat analogous to the exponential suppression 
encountered in tunneling phenomena in quantum mechanics. This 
suppression of monopole production in Drell-Yan production does 
not depend on whether the semi-classical monopole solution is 
spherically symmetric or not. More recently, it has been 
suggested that this challenge may be overcome at colliders by 
exploiting the magnetic analogue of the Schwinger mechanism. In 
the presence of adequately strong magnetic fields the (dual) 
Schwinger mechanism may lead to an observable cross section for 
monopole pair production in heavy ion collisions -- for a recent 
discussion and additional references, see Ref.~\cite{Schwinger}. 
It is fair to state that the production mechanisms in colliders 
of more complex topological structures such as necklaces requires 
additional studies well beyond the scope of this paper.

\section{Strings and Necklaces}
\label{sec:Neck}

Around the string that connects the $SU(3)_L$ and $SU(3)_R$ 
monopoles, $\vev{h_u}$ remains constant implying that there 
are no transverse zero modes in the up-type quark sector. 
However, the phases of $\vev{h_d}$ and $\vev{N}$ change by 
$-2\pi$ and $2\pi$ respectively. The masses of the down-type 
quarks can be written as 
\beq
\mathcal{M}_d=
\begin{pmatrix}
g^c, & d^c
\end{pmatrix}
\begin{pmatrix} 
\vev{N}, & 0\\
& \\
\vev{\nu^c}, & \vev{h_d}
\end{pmatrix}
\begin{pmatrix}
g \\
&\\
d
\end{pmatrix}.
\label{matrix}
\eeq      
Three of the four $3\times 3$ blocks in the mass matrix are of 
the order of $\vev{N}$, $\vev{\nu^c}$, and $\vev{h_d}$ as 
indicated 
with constant unsuppressed coefficients. The fourth block is 
suppressed by powers of the Planck mass since the relevant 
direct trilinear Yukawa coupling is forbidden by $E_6$. 
Applying the results of Ref.~\cite{ganoulis}, we then see that 
there exist nine right-moving and nine left-moving zero modes 
(one for each family and color). A very similar analysis can be 
done for the charged leptons. We conclude that these strings 
are superconducting. In contrast, the string that connects the 
$SU(2)_R$ and $SU(3)_L$ monopoles is not superconducting since 
$\vev{N}$, $\vev{h_u}$, and $\vev{h_d}$ remain constant as we go 
around it. It is worth mentioning that the fact that the phase of
$\vev{\nu^c}$ changes by $-2\pi$ around this string does not 
imply the existence of zero modes in this case. In order to see 
this, we employ a theorem given in Ref.~\cite{ganoulis} which 
states that, if a particular mass matrix element remains constant 
around the string, we can remove from the mass matrix the row 
and the column that contain it when calculating the number of 
transverse zero modes. In our case $\vev{N}$ and $\vev{h_d}$ 
remain unaltered around the string, so all rows and columns can 
be removed and no zero modes appear.

\begin{figure}[t]
\centerline{\epsfig{file=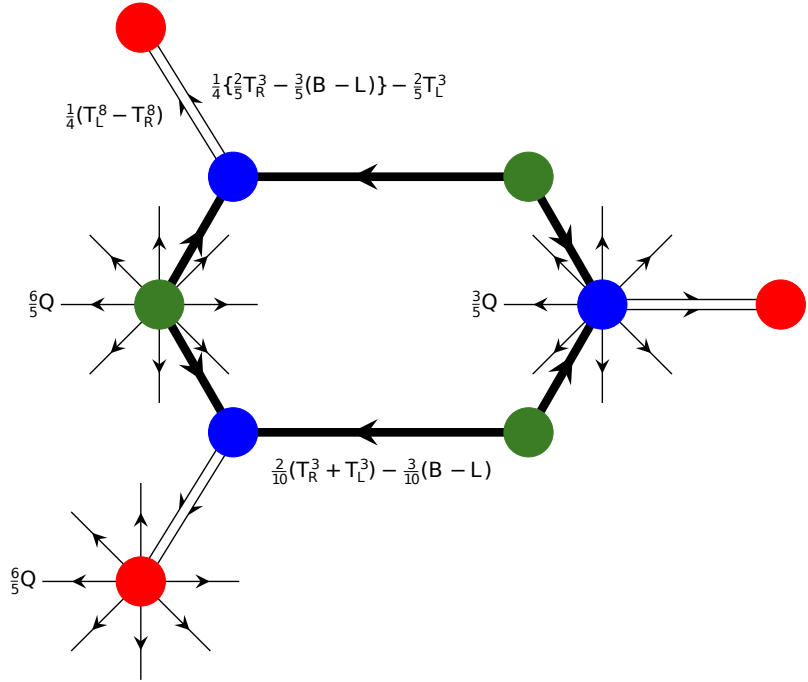,width=10.6cm}}
\caption{Necklace configuration with alternating $SU(3)_L$ 
(blue) and $SU(2)_R$ (green) monopoles from the symmetry 
breaking $G
%=SU(3)_c\times SU(3)_L\times SU(3)_R
\to SU(3)_c\times 
SU(2)_L\times U(1)_{Y_L}\times U(1)_{Y_R}\times U(1)_R
\to SU(3)_c\times SU(2)_L\times U(1)_Y\times Z_2\to 
SU(3)_c\times U(1)_{em}\times Z_2$. These are connected by 
half flux tubes along the necklace as indicated. Each 
$SU(3)_L$ (blue) monopole in the necklace is also connected 
by a flux tube with an $SU(3)_R$ (red) monopole hanging 
outside the necklace. We display explicitly only the Coulomb 
magnetic flux of three of the constituent monopoles and the 
flux along two of the tubes.}
\label{fig:FangNeck}
\end{figure}

Let us now turn to the alternative case where the symmetry 
breaking of $E_6$ employs the $\nu^c\nu^c$ component of a Higgs 
$\overline{{\bf 351}'}$. In this case, intermediate scale $Z_2$ 
topologically stable strings are produced \cite{TopDef,z2string} 
in addition to the superheavy Dirac and the intermediate scale 
triply charged monopoles. A rotation by $2\pi/10$ around the 
generator in Eq.~(\ref{orthgen1}) is now left unbroken by the 
VEV of $\nu^c\nu^c$ since its relevant charges are $T^3_R=-2$, 
$B-L=2$. Consequently, the flux tube from the $SU(2)_R$ to 
$SU(3)_L$ monopole splits into two equivalent tubes with 
magnetic flux 
\beq
\frac{2}{10}T^3_R-\frac{3}{10}(B-L).
\label{halftube}
\eeq 
After the electroweak symmetry breaking, this tube acquires an 
extra magnetic flux $T^3_L/5$ so that $\vev{h_u}$, $\vev{h_d}$ 
remain constant around it. One can show that this ``half flux 
tube'' is not superconducting. The combined flux tube though is 
not affected. We can imagine that we break one of the 
two strings from the $SU(2)_R$ to $SU(3)_L$ monopole which leaves 
the two monopoles connected by one string and two ``loose'' 
strings attached to the two monopoles. One can then connect 
these latter strings to other similar monopole-string structures 
in series to form ``fang necklaces'' -- see 
Fig.~\ref{fig:FangNeck}. More complex fang necklaces 
can be contemplated where each $SU(3)_L$ monopole (antimonopole) 
in the necklace is connected by a half tube either to its own 
antimonopole or an $SU(2)_R$ monopole (antimonopole), and each 
$SU(2)_R$ monopole (antimonopole) either to its own antimonopole 
or to an $SU(3)_L$ monopole (antimonopole). Each $SU(3)_R$ monopole 
(antimonopole) in the necklace is also connected by a flux tube 
to an $SU(3)_R$ monopole (antimonopole) hanging outside the 
necklace or to its own antimonopole which participates in a 
different necklace.   

\section{Conclusions} 
\label{sec:concl}

The trinification group $SU(3)_c\times SU(3)_L\times SU(3)_L$ 
implements charge quantization and predicts the existence of a 
topologically stable monopole of magnetic charge $6\pi/e$. The 
trinification symmetry breaking to the SM may occur 
in a number of steps, and we have discussed a scenario in which 
this monopole may be regarded as a magnetic baryon, in rough 
analogy with the QCD baryon. It is composed 
of three confined monopoles (magnetic quarks), where the latter 
monopoles carry some Coulomb magnetic flux accompanied by a 
magnetic flux tube. These confined monopoles can yield more 
elaborate topological configurations and we display one such 
example consisting of a fang necklace. In contrast to the 
superheavy GUT monopoles the trinification monopole discussed 
here may be accessible at high energy colliders.

\section*{Acknowledgments}
This work is supported by the Hellenic Foundation for Research 
and Innovation (H.F.R.I.) under the ``First Call for H.F.R.I. 
Research Projects to support Faculty Members and Researchers and 
the procurement of high-cost research equipment grant'' (Project 
Number: 2251). Q.S. is supported in part by the DOE Grant 
DE-SC-001380. We thank Joey Betz for preparing the figures for 
this paper.

\def\ijmp#1#2#3{{Int. Jour. Mod. Phys.}
{\bf #1},~#3~(#2)}
\def\plb#1#2#3{{Phys. Lett. B }{\bf #1},~#3~(#2)}
\def\zpc#1#2#3{{Z. Phys. C }{\bf #1},~#3~(#2)}
\def\prl#1#2#3{{Phys. Rev. Lett.}
{\bf #1},~#3~(#2)}
\def\rmp#1#2#3{{Rev. Mod. Phys.}
{\bf #1},~#3~(#2)}
\def\prep#1#2#3{{Phys. Rep. }{\bf #1},~#3~(#2)}
\def\prd#1#2#3{{Phys. Rev. D }{\bf #1},~#3~(#2)}
\def\npb#1#2#3{{Nucl. Phys. }{\bf B#1},~#3~(#2)}
\def\np#1#2#3{{Nucl. Phys. B }{\bf #1},~#3~(#2)}
\def\npps#1#2#3{{Nucl. Phys. B (Proc. Sup.)}
{\bf #1},~#3~(#2)}
\def\mpl#1#2#3{{Mod. Phys. Lett.}
{\bf #1},~#3~(#2)}
\def\arnps#1#2#3{{Annu. Rev. Nucl. Part. Sci.}
{\bf #1},~#3~(#2)}
\def\sjnp#1#2#3{{Sov. J. Nucl. Phys.}
{\bf #1},~#3~(#2)}
\def\jetp#1#2#3{{JETP Lett. }{\bf #1},~#3~(#2)}
\def\app#1#2#3{{Acta Phys. Polon.}
{\bf #1},~#3~(#2)}
\def\rnc#1#2#3{{Riv. Nuovo Cim.}
{\bf #1},~#3~(#2)}
\def\ap#1#2#3{{Ann. Phys. }{\bf #1},~#3~(#2)}
\def\ptp#1#2#3{{Prog. Theor. Phys.}
{\bf #1},~#3~(#2)}
\def\apjl#1#2#3{{Astrophys. J. Lett.}
{\bf #1},~#3~(#2)}
\def\apjs#1#2#3{{Astrophys. J. Suppl.}
{\bf #1},~#3~(#2)}
\def\n#1#2#3{{Nature }{\bf #1},~#3~(#2)}
\def\apj#1#2#3{{Astrophys. J.}
{\bf #1},~#3~(#2)}
\def\anj#1#2#3{{Astron. J. }{\bf #1},~#3~(#2)}
\def\mnras#1#2#3{{MNRAS }{\bf #1},~#3~(#2)}
\def\grg#1#2#3{{Gen. Rel. Grav.}
{\bf #1},~#3~(#2)}
\def\s#1#2#3{{Science }{\bf #1},~#3~(#2)}
\def\baas#1#2#3{{Bull. Am. Astron. Soc.}
{\bf #1},~#3~(#2)}
\def\ibid#1#2#3{{\it ibid. }{\bf #1},~#3~(#2)}
\def\cpc#1#2#3{{Comput. Phys. Commun.}
{\bf #1},~#3~(#2)}
\def\astp#1#2#3{{Astropart. Phys.}
{\bf #1},~#3~(#2)}
\def\epjc#1#2#3{{Eur. Phys. J. C}
{\bf #1},~#3~(#2)}
\def\nima#1#2#3{{Nucl. Instrum. Meth. A}
{\bf #1},~#3~(#2)}
\def\jhep#1#2#3{{J. High Energy Phys.}
{\bf #1},~#3~(#2)}
\def\jcap#1#2#3{{J. Cosmol. Astropart. Phys.}
{\bf #1},~#3~(#2)}
\def\lnp#1#2#3{{Lect. Notes Phys.}
{\bf #1},~#3~(#2)}
\def\jpcs#1#2#3{{J. Phys. Conf. Ser.}
{\bf #1},~#3~(#2)}
\def\aap#1#2#3{{Astron. Astrophys.}
{\bf #1},~#3~(#2)}
\def\mpla#1#2#3{{Mod. Phys. Lett. A}
{\bf #1},~#3~(#2)}


\begin{thebibliography}{}

\bibitem{GeorgiGlashow}
H.~Georgi and S.L.~Glashow, Phys. Rev. Lett. {\bf 32}, 
438 (1974). 
%%CITATION = doi:10.1103/PhysRevLett.32.438;%%

\bibitem{dokos}
C.P. Dokos and T.N.~Tomaras, Phys. Rev. D {\bf 21}, 
2940 (1980).
%%CITATION =  doi:10.1103/PhysRevD.21.2940;%%

\bibitem{daniel}
M.~Daniel, G.~Lazarides, and Q.~Shafi, Nucl. Phys. 
{\bf B170}, 156 (1980); 
%%CITATION = NUPHA,B170,156;%%
G.~Lazarides, Q.~Shafi, and W.P.~Trower, Phys. Rev. 
Lett. {\bf 49}, 1756 (1982).
%%CITATION = PRLTA,49,1756;%%

\bibitem{monopole}
G.~'t~Hooft, Nucl. Phys. {\bf B79}, 276 (1974);
%%CITATION = NUPHA,B79,276;%%"
A.M. Polyakov, J. Exp. Theor. Phys. Lett. {\bf 20}, 
194 (1974).
%%CITATION = JTPLA,20,194;%%

\bibitem{pati} 
J.C.~Pati and A.~Salam,
Phys. Rev. D {\bf 10}, 275 (1974),
Erratum: Phys. Rev. D {\bf 11}, 703 (1975).
%%CITATION = PHRVA,D10,275;%%

\bibitem{magg}
G.~Lazarides, M.~Magg, and Q.~Shafi,
Phys. Lett. {\bf 97B}, 87 (1980).
%%CITATION = PHLTA,97B,87;%%

\bibitem{TopDef}
G.~Lazarides and Q. Shafi, J. High Energy Phys. {\bf 10}, 
193 (2019).
%%CITATION = doi:10.1007/JHEP10(2019)193;%%

\bibitem{volovik}
G.E.~Volovik, J. Exp. Theor. Phys. {\bf 131}, 11 (2020), 
Sov. Phys. JETP {\bf 131}, 11 (2020).
%%CITATION = doi:10.1134/S1063776120070146;%%

\bibitem{kibble}
T.W.B.~Kibble, G.~Lazarides, and Q.~Shafi, Phys. Rev. D 
{\bf 26}, 435 (1982).
%%CITATION = doi:10.1103/PhysRevD.26.435;%%

\bibitem{volovik2019}
J.T.~M\"{a}kinen, V.V.~Dmitriev, J.~Nissinen, J.~Rysti, 
G.E.~Volovik, A.N.~Yudin, K.~Zhang, and V.B.~Eltsov, Nature 
Commun. {\bf 10}, 237 (2019);
%%CITATION = ARXIV:1807.04328;%%
G.E.~Volovik and K.~Zhang, Phys. Rev. Res. {\bf 2}, 023263 
(2020);
%%CITATION = doi:10.1103/PhysRevResearch.2.023263;%%
K.~Zhang, Phys. Rev. Res. {\bf 2}, 043356 (2020). 
%%CITATION = doi:10.1103/PhysRevResearch.2.043356;%%

\bibitem{kephart}
Q.~Shafi and C. Wetterich, NATO Sci. Ser. B {\bf 111}, 47 
(1984);
%%CITATION = BA-83-45;%%
G.~Lazarides, C.~Panagiotakopoulos, and Q.~Shafi, Phys. Rev. Lett. 
{\bf 58}, 1707 (1987);
%%CITATION = PRLTA,58,1707;%%
G.~Lazarides, Q.~Shafi, and T.N.~Tomaras, Phys. Rev. D {\bf 39}, 
1239 (1989);
%%CITATION = PHRVA,D39,1239;%%
T.W.~Kephart, C.-A.~Lee, and Q.~Shafi, J. High Energy 
Phys. {\bf 01}, 088 (2007);
%%CITATION = HEP-PH/0602055;%%
T.W.~Kephart, G.K.~Leontaris, and Q.~Shafi, J. High Energy 
Phys. {\bf 10}, 176 (2017).
%%CITATION = ARXIV:1707.08067;%%
 
\bibitem{ramond} 
F.~G\"{u}rsey, P.~Ramond, and P.~Sikivie,
Phys. Lett. {\bf 60B}, 177 (1976);
%%CITATION = PHLTA,60B,177;%%"
Y.~Achiman and B.~Stech,
Phys. Lett. {\bf 77B}, 389 (1978);
%%CITATION = PHLTA,77B,389;%%"
Q.~Shafi, Phys. Lett. {\bf 79B}, 301 (1978).
%%CITATION = PHLTA,79B,301;%%"

\bibitem{dirac}
P.A.M.~Dirac, Proc. Roy. Soc. Lond. A {\bf 133}, no. 821, 
60 (1931).
%%CITATION = doi:10.1098/rspa.1931.0130;%%

\bibitem{WenWitten}
X.-G.~Wen and E.~Witten, Nucl. Phys. {\bf B261}, 651 
(1985).
%%CITATION = doi:10.1016/0550-3213(85)90592-9;%%

\bibitem{Ztube}
Y.~Nambu, Phys. Rev. D {\bf 10}, 4262 (1974);
%%CITATION = PHRVA,D10,4262;%%
Y.~Nambu, Nucl. Phys. {\bf B130}, 505 (1977);
%%CITATION = NUPHA,B130,505;%%
G.~Lazarides and Q.~Shafi, Phys. Lett. {\bf 94B}, 149 
(1980).
%%CITATION = doi:10.1016/0370-2693(80)90845-X;%%

\bibitem{hung}
P.Q.~Hung, Nucl. Phys. {\bf B962}, 115278 (2021).
%%CITATION = doi:10.1016/j.nuclphysb.2020.115278;%%

\bibitem{moedal}
B.~Acharya et al. [MoEDAL Collaboration], J. High Energy 
Phys. {\bf 08}, 067 (2016);
%%CITATION = doi:10.1007/JHEP08(2016)067;%%
G.~Aad et al. [ATLAS Collaboration], Phys. Rev. Lett. 
{\bf 124}, 031802 (2020);
%%CITATION = doi:10.1103/PhysRevLett.124.031802;%%
M.M.H.~El Sawy, Ph.D. Thesis, preprint CERN-THESIS-2020-191,
{\tt https://cds.cern.ch/record/2744867}.

\bibitem{DrellYan}
N.E.~Mavromatos and V.A.~Mitsou, Int. J. Mod. Phys. A {\bf 35}, 
203001 (2020).
%%CITATION = doi:10.1142/S0217751X20300124;%%

\bibitem{Schwinger}
O.~Gould, D.L.-J. Ho, and A.~Rajantie, arXiv:2103.14454.
%%CITATION = ARXIV:2103.14454;%%

\bibitem{ganoulis} 	
N.~Ganoulis and G.~Lazarides, Phys. Rev. D {\bf 38}, 547 (1988);
%%CITATION = doi:10.1103/PhysRevD.38.547;%%
N.~Ganoulis and G.~Lazarides, Nucl. Phys. {\bf B316}, 443 (1989).
%%CITATION = NUPHA,B316,443;%%

\bibitem{z2string}
T.W.B.~Kibble, G.~ Lazarides, and Q.~Shafi, 
Phys. Lett. {\bf 113B}, 237 (1982).
%%CITATION = PHLTA,113B,237;%%

\end{thebibliography}
\end{document}